*Superhydrophobic frictions*

*Timothée Mouterde\*, Pascal Raux\*, Christophe Clanet & David Quéré*

(\*) These authors contributed equally to this work.

Physique et Mécanique des Milieux Hétérogènes,

UMR 7636 du CNRS, ESPCI, PSL Research University, Paris, France.

and

LadHyX, UMR 7646 du CNRS, École polytechnique, 91128 Palaiseau Cedex, France.

Contrasting with its sluggish behavior on standard solids, water is extremely mobile on superhydrophobic materials, as shown for instance by the continuous acceleration of drops on tilted water-repellent leaves. For much longer substrates, however, drops reach a terminal velocity that results from a balance between weight and friction, allowing us to question the nature of this friction. We report that the relationship between force and terminal velocity is non-linear. This is interpreted by showing that classical sources of friction are minimized, so that the aerodynamical resistance to motion becomes dominant, which eventually explains the matchless mobility of water. Our results are finally extended to viscous liquids, also known to be unusually quick on these materials.

Despite its low viscosity, water running down tilted solids is surprisingly lazy. Its worm's pace arises from contact line, whose presence induces pining and magnifies viscous dissipation, which contributes to slow down and even stop the liquid [1-2]. In contrast, water on superhydrophobic (SH) materials move at unrivalled speeds, owing to the conjunction of minimized pining and maximized contact angle. While drops on tilted plastic or glass immediately reach a velocity of typically 1 cm/s [3-4], water on non-wetting materials speeds up by decimeter-size or meter-size distances [5-8] before reaching a terminal speed $U$ as high as a few meters per second. In such Galileo-like experiments, the drop is subjected to an acceleration $g\sin\alpha$, denoting $g$ the acceleration of gravity and $\alpha$ the tilting angle, possibly diminished by the (weak) pining on the solid [9-15]. The stationary regime of descent is observed when the weight is balanced by the friction acting on the moving drop, a resistance that remains to be characterized on superhydrophobic materials. Our aim in this paper is to deduce the nature of this friction from direct measurements, contrasting with previous studies performed in transient regimes [12-13] or inside rotating SH cylinders [16]. In all the latter studies, friction was assumed to be simply viscous, which we question in this paper.

The substrates in our experiments are long brass bars rendered water-repellent by a spray of colloidal suspension of hydrophobic silica nanobeads in isopropanol (Glaco Mirror Coat Zero; Soft99). The resulting texture imaged by atomic force microscopy is shown in Figures 1a and 1b both in the plane of the material and perpendicular to it. The surface exhibits cavities and bumps at the scale of 100 nm and the root mean square roughness (RMS) deduced from AFM pictures is $35 \pm 5$ nm, which is comparable to the mean size of the silica nanobeads and to the average top-to-bottom distance in Figure 1b. This simple and reproducible treatment allows us to coat long solids (around 2.5 m), a necessary condition for reaching the terminal velocity $U$ of water drops (Figure S1 in the Supplemental Information). Starting with an initial acceleration $g\sin\alpha$, a typical distance of $U^2/2g\sin\alpha$ is needed to reach $U$, that is for $U=1$ m/s and $\sin\alpha=0.1$, about 1 m – a length significantly smaller than that of our inclines.

Advancing and receding angles of water are $\theta_a = 171 \pm 2°$ and $\theta_r = 165 \pm 2°$, with the high angles and low hysteresis $\Delta\theta = \theta_a - \theta_r$ typical of SH materials [17]. Due to the hydrophobic nature of the coating, water remains upon the roughness and contacts a mixture of solid and air, as evidenced by the silvery aspect of its base. Using Cassie formula, we can deduce from contact angles the proportion $\phi$ of solid/water contact, and find $\phi \approx 3.5 \pm 2.0$ %, a value much smaller than unity. Aging of our materials is quite slow, and the quality of non-wetting is regularly controlled. If degraded or damaged, the surface is simply regenerated by a new treatment.

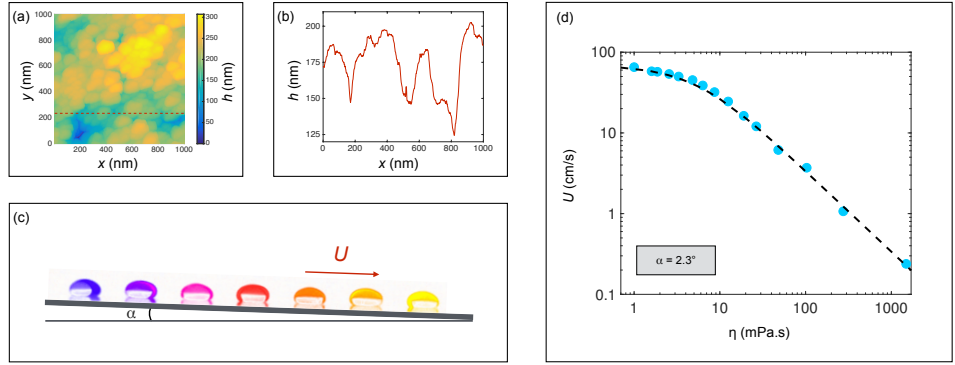

**Figure 1.** Drops running down superhydrophobic materials. **(a)** AFM picture of the material used in our experiments: $x$ and $y$ are the coordinates in the plane and $h$ is the depth. **(b)** AFM cross-sectional view along the red dotted line in (a). **(c)** Water drop descending a tilted SH-plate in the regime where it reached its terminal velocity $U = 66$ cm/s. We superimpose colored images separated by 20 ms for a volume $\Omega = 100$ µL, an equatorial radius $R \approx 3.7$ mm and a tilt $\alpha = 2°$. (See also Movie 1.) **(d)** Terminal velocity $U$ of water-glycerol mixtures ($\Omega = 100$ µL) as a function of their viscosity $\eta$. The tilt angle here is $\alpha = 2.3°$ and the dashed line represents Eq. 3.

An example of experiment is shown in Figure 1c, where we superimpose colored images of a water drop (volume $\Omega = 100$ µL, surface tension $\gamma = 72$ mN/m, viscosity $\eta = 1$ mPa.s and density $\rho = 1000$ kg/m$^3$) running down a SH plate tilted by 2°. This figure is extracted from a high-speed movie shot at 2000 frames per second. We first notice that the drop has reached its terminal velocity (here $U = 66$ cm/s). In addition, despite its high mobility, the drop keeps a quasi-static shape: it is slightly flattened by gravity as expected from its size, above the capillary length $a = (\gamma/\rho g)^{1/2}$ ($a^3 \approx 20$ µL), and its front-rear symmetry evidences the small value of the hysteresis $\Delta\theta$. Denoting $R$ as the equatorial drop radius, the adhesion force opposing the motion scales as $\gamma R (\cos\theta_r - \cos\theta_a)$, which reduces to $\gamma R \sin\theta \, \Delta\theta$ at small $\Delta\theta$ [3, 5]. For $\Omega = 100$ µL, adhesion of water on our materials ($\theta \approx 168°$, $\Delta\theta \approx 6°$) is small compared to the gravity force $\rho g \Omega \sin\alpha$ for $\alpha > 0.2°$. Thus, the drop terminal speed directly results from a balance between the projected weight $\rho g \Omega \sin\alpha$ and the (unknown) friction force $F(U)$.

The aim of this paper is to characterize the function $F(U)$ and consequently to understand what fixes the terminal velocity of drops. Contrasting with the case of partial wetting, where friction is dominated by viscous effects around the contact line [2], the friction on superhydrophobic materials was up to now assumed to arise from viscous effects inside the liquid [12-16]. We first check the influence of viscosity η by using water/glycerol mixtures, which provides variations from η = 1 mPa.s to η = 1490 mPa.s. Figure 1d presents the terminal speed $U$ of drops as a function of η, for a tilt α equal to 2.3°. At large η, $U$ strongly decreases as the liquid gets more viscous, which highlights the dominant role of viscosity in the resistance to motion. In contrast, viscous effects for η < 10 mPa.s (including the important case of water) tend to become marginal, as seen from the tendency of the data to plateau. In what follows, we discuss these two successive regimes of friction.

Focusing first on the viscous case, we plot in Figure 2a the terminal speed $U$ of drops as a function of the slope sin α, for η = 110 mPa.s and two volumes (Ω = 100 μL and 200 μL).

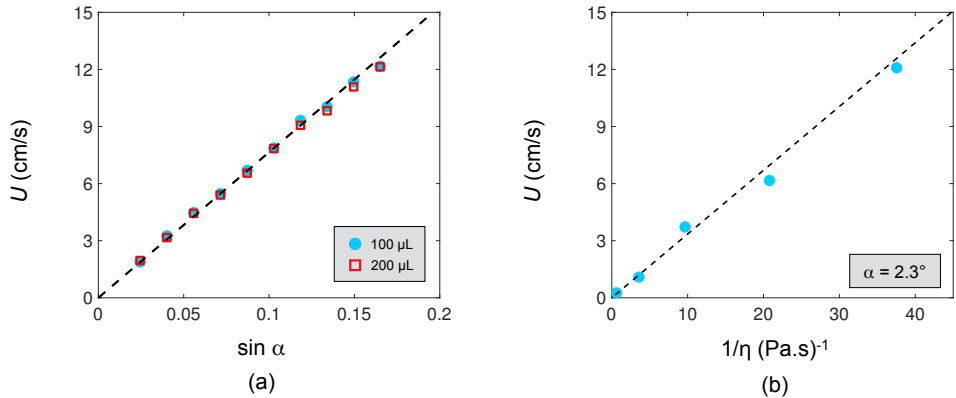

**Figure 2.** Mobility of viscous drops on SH materials tilted by an angle α. **(a)** Terminal velocity $U$ as a function of the sine of α for a water-glycerol mixture with viscosity η ≈ 110 mPa.s and Ω ≈ 100 μL (blue circles) or Ω ≈ 200 μL (red squares). (See also Movie 2.) **(b)** Terminal velocity $U$ of viscous drops as a function of 1/η, the inverse of the viscosity, for α = 2.3° and Ω ≈ 100 μL. The linear fit (dashed line) has a slope of 3.35 mN/m.

Figure 2a shows that speed is linear in slope sin α, as expected in a viscous regime [18-20]. Indeed, denoting δ as the characteristic scale of velocity gradients, the viscous force varies as $\eta U\Sigma/\delta$, denoting Σ as the surface area of the drop base. For puddles, the volume Ω and surface area Σ are simply proportional to each other (Ω ~ Σa ~ $\pi R^2 a$), and velocity gradients develop across the thickness of the puddles (δ ~ a). Hence the balance of the viscous force with the weight ρgaΣsin α yields:

$$U \sim \frac{\gamma}{\eta}\sin\alpha \quad (1)$$

The velocity is expected to be linear in $\sin\alpha$ and independent of the volume $\Omega$, as observed in Figure 2a and also in Figure S2. Eq. (1) also predicts that $U$ decreases hyperbolically with the viscosity $\eta$, which we check in Figure 2b for $\alpha = 2.3°$. $U$ is observed to increase linearly with $1/\eta$, and the slope deduced from the fit (dashed line in the figure), 3.35 mN/m, nicely compares to $\gamma\sin\alpha \approx 2.9$ mN/m. This allows us to evaluate a prefactor of 0.85 in the scaling formula of the force, which finally writes $F \approx 0.85\, \eta U\Omega/a^2$.

As seen in Figure 1d, a purely viscous friction does not describe the behavior at small $\eta$. These large deviations are confirmed by plotting the terminal velocity of water drops ($\eta = 1$ mPa.s) as a function of the slope $\sin\alpha$ (Figure 3a). Instead of the linear behavior reported in Figure 2a, we now observe a concave curve that highlights the existence of a supplementary friction.

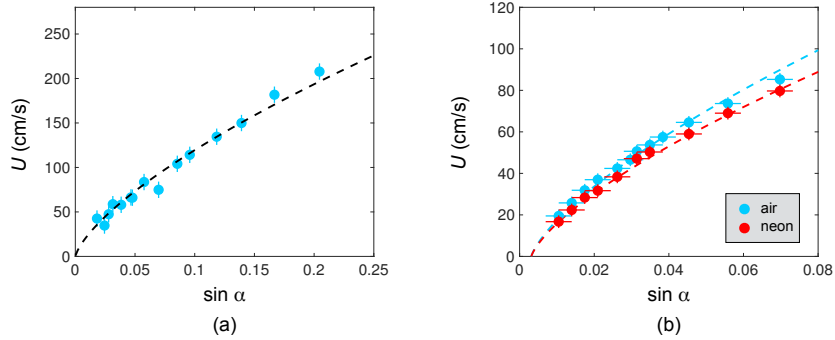

**Figure 3.** Water drop mobility on SH inclines. In all plots, the drop volume is $\Omega = 100$ µL and dashed lines show Eq. 3. **(a)** Terminal velocity $U$ as a function of the sine of the sliding angle $\alpha$ for water drops. **(b)** Terminal velocity $U$ as a function of $\sin\alpha$ for water drops in an atmosphere of air (light blue circles) or neon (red circles).

Water runs down at velocities of typically 1 m/s, comparable to that of raindrops, which suggests an aerodynamical drag. The Reynolds number in air $Re = 2\rho_a RU/\eta_a$ (defined with the diameter $2R$ of the drop and with the air viscosity $\eta_a = 18$ µPa.s and density $\rho_a = 1$ kg/m³) is typically between 100 and 1000. The ratio between viscous friction in water (that scales as $\eta U\Sigma/a$) and inertial friction in air (that scales as $\rho_a U^2\Sigma$) is $\rho_a Ua/\eta$, a quantity independent of the drop volume. For water, this number is of order unity, confirming the aerodynamical origin of the additional friction. More generally, the drag force in air can be written:

$$F \approx \rho_a C_d(Re) S U^2/2 \qquad (2)$$

where $S$ is a surface area and the drag coefficient $C_d$ is a function of $Re$. In our Reynolds range, the drag results from a skin friction, that is, the friction at stake in the air boundary layer around the drop. This statement has two consequences: (1) The relevant surface area $S$ in equation 2 is the top surface of the drop, whose area scales as $\Sigma$. (2) The drag coefficient is $C_d(Re) = 2/Re^{1/2}$, as deduced by balancing equation 2 with the viscous air friction in the boundary layer whose thickness is classically given by $\delta \sim (\eta_a R/\rho_a U)^{1/2} \sim R/Re^{1/2}$. Adding viscous and aerodynamical frictions, we get an equation for the terminal velocity of the drop:

$$\rho g a \sin\alpha \approx x\, \eta U/a + y\, \rho_a U^2/Re^{1/2} \qquad (3)$$

where the first prefactor $x$ can be extracted from Figure 2 ($x \approx 0.85$) and where the second prefactor $y$ has to be determined. As seen in Figure 1d, a unique value of $y$ ($y \approx 34$, dashed line) allows us to adjust data at all viscosities, from pure water to pure glycerol. The high value of $y$ can be explained by the fact that non-wetting drops rotate as they move (see figure S3), which is known for rolling spheres to increase $C_d$ by a factor ~10 compared to sliding spheres [21]. Moreover, assuming a disk-shape for the drop, we may underestimate the top surface $S$ where the boundary layer develops. The adjustment of our data by equation 3 seems robust: using the same prefactors, we can deduce the relationship between velocity and slope and there again the model nicely adjusts the data with water (Figure 3a). The term varying as $U^{3/2}$ in equation 3 dominates over the viscous linear term at large velocity, so that we expect $U$ to asymptotically vary as $\sin^{2/3}\alpha$, a simple way to explain the concavity of the curve in Figure 3a – while the viscous term dictates a linear behavior at small tilt.

As an implication of equation 3, we expect the drop velocity to depend on the nature of the surrounding air. It is challenging to test this dependency, but we tried it by using neon, whose viscosity $\eta_a = 29$ µPa.s and density $\rho_a = 0.9$ kg/m$^3$ make its kinematic viscosity $\eta_a/\rho_a$ larger by 60% than that of air. We performed the experiment described in Figure 1c in a closed transparent box filled either with air or with neon. Figure 3b shows our results for water drops with $\Omega = 100$ µL, the terminal velocity $U$ being plotted as a function of the slope $\sin\alpha$. Owing to the box size, the experimental range of accessible slopes to reach terminal velocity is limited, which reveals the small offset invisible in Figure 3a and due to the residual adhesion of water (drops move if the substrate is tilted by more than 0.3°). However, we clearly distinguish data obtained in air from that in neon: drops systematically move slower in the latter case, showing the existence of a larger friction, in agreement with eq. 3 where an increase of the kinematic viscosity of the surrounding gas should induce such an effect. The colored dashes in the figure show the predictions of equation 3 with the parameters of the respective gas and the same prefactors as previously ($x = 0.85$, $y = 34$). The fits are found to be satisfactory, showing in particular the same weak, yet significant, differences between the two systems.

Water moving along superhydrophobic materials develop original dynamical behaviors, compared to the usual cases where the drop velocity is rather fixed by viscous effects close to the contact line. For water-repellent materials, such effects are negligible and viscosity is found to be only a weak correction to aerodynamical effects. The ratio between viscous and aerodynamical forces in equation 3, $y\rho_a Ua/\eta Re^{1/2}$, is typically 4 in our experiments (considering that $x \sim 1$), implying that aerodynamical drag is the main force opposing the drop motion – an original situation in wetting dynamics. Assuming it is dominant, we can deduce the plateau velocity in Figure 1d (corresponding to "large" slopes or small viscosities) by simply balancing the weight with the aerodynamical force in equation 3:

$$U \sim \left(\frac{\gamma R \rho g \sin^2 \alpha}{\eta_a \rho_a y^2}\right)^{1/3} \qquad (4)$$

This speed is independent of the liquid viscosity and expected to be ~2 m/s for a tilt angle of 10°, in good agreement with observations. Such a high velocity shows that the mobility of water is increased by a factor of at least 100 on repellent materials compared to usual ones – reflecting the conjunction of highly reduced adhesion and highly reduced friction, both arising from the air trapped in the texture. About this thin layer, we can wonder whether its presence could also generate a significant (and specific) friction. Such a friction is promoted by the thinness of the air film and by the existence of a slip at the water/air interface, at the drop base. As seen in Figures 1b and 1c, the roughness has comparable wavelength and height $h$, which sets a slip length of order $h$ [22-23]. This means that the slip velocity $U_s$ scales as $Uh/a$, and thus that the stress in the air film is of order $\eta_a U_s/h \sim \eta_a U/a$. If this stress were balanced by the drop weight, the resulting velocity would depend on the air viscosity, in qualitative agreement with Figure 3b, but be proportional to the slope, in strong disagreement with Figure 3a. More fundamentally, this additional friction is found to be smaller by a factor $\eta/\eta_a \sim 50$ than the viscous force in water and typically 100 times smaller than the aerodynamical force in equation 3, showing that the air film, so crucial in superhydrophobic states, has a negligible impact on the friction.

Drops on SH materials are highly mobile and what limits their mobility depends on their viscosity. Viscous liquids are slowed by the dissipation in the bulk, a consequence of the rolling motion accompanying the drop translation. In contrast, the internal dissipation in water, of low viscosity, is weaker than the aerodynamical resistance. This minimizes the role of the substrate, whose influence can be however revealed by a closer view. (1) A solid substrate generates a small (yet measurable) contact angle hysteresis, which can stop the liquid at small tilt and slightly lower the speed at larger tilt. (2) Less trivially, the drag coefficient needed to fit the results is significantly larger than that for a raindrop in pure translation, as also observed for rotating objects along planes [21]. Indeed, even if viscosity is only a small correction to aerodynamical effects, it induces rotation in water (Figure S3 and Movie 3), which in turn increases the drag coefficient. (3) In the same vein, even a marginal viscous friction due to the no-slip boundary condition at the substrate can impact the drop shape. We considered here situations where this shape remains quasi-static. However, at higher substrate tilt, that is, at higher drop velocity $U$, the capillary number $Ca = \eta U/\gamma$ can become large enough to imply changes in the drop shape. Water subjected to a viscous force will elongate, which is indeed what we find when substrates are inclined by ~20° or more. The critical capillary number at which such deformations are found is on the order of $10^{-2}$, a relevant value for a dynamical wetting transition [24] – even if this problem remains to be discussed in non-wetting situations. The use of repellent materials in this limit remains highly valuable, since we do not observe any continuous deposition, owing to the high speed of dewetting on SH materials. Of course, drop deformation might in turn impact the friction law, which remains to be described. More generally, the universality of our model (*i.e.* the fact that the detail of the solid texture does not seem to matter in Eq. 3) should be explored in the future, when technology will allow us to microfabricate the long, controlled substrates needed for such studies. The case of smaller water drops would also deserve a dedicated study. First, we expect their motion to be highly sensitive to the hysteretic adhesion, that now writes $(\gamma R^2/a)\sin\theta\, \Delta\theta$, and thus can become comparable to the weight $\rho R^3 g\sin\alpha$ at small radius $R$. Yet, even in an ideal situation without hysteresis, we anticipate a modification of both viscous and aerodynamical frictions, due to the spherical shape. The rolling motion of a sphere minimizes its internal dissipation [18], while the air skin drag friction now applies on a typical surface area $R^2$. Hence, in the case of a dominant air friction, the terminal velocity $U$ should scale as $R\,(\rho^2 g^2 \sin^2\alpha/\rho_a\eta_a)^{1/3}$, an expression quite different from that for larger drops (Eq. 4) – and remarkably, more sensitive to gravity, despite the size reduction!

Author contributions: T.M., P.R. and D.Q. conceived and designed the research. T.M. and P.R. performed the experiments. T.M., P.R., C.C. and D.Q. analyzed the data, elaborated the models and wrote the manuscript.

The authors declare no conflict of interest.

Acknowledgements: We thank Philippe Bourrianne, Antoine Fosset and Evan Spruijt for their help in the experiments, and Emmanuel du Pontavice for valuable discussions.

*Superhydrophobic frictions*

Timothée Mouterde*, Pascal Raux*, Christophe Clanet & David Quéré

(*) These authors contributed equally to this work.

Physique et Mécanique des Milieux Hétérogènes,

UMR 7636 du CNRS, ESPCI, PSL Research University, Paris, France.

and

Ladhyx, UMR 7646 du CNRS, École Polytechnique, 91128 Palaiseau Cedex, France.

**Supplemental information**

We present in these supplementary materials additional experiments and follow the development of the accompanying paper. 1) We show the terminal velocity attained by both water and viscous drops. 2) We discuss the influence of the volume on the superhydrophobic friction in the viscous and inviscid cases. 3) We show experimental evidences of liquid and gas dissipation. 4) We provide captions for the movies.

**1. Terminal velocity**

Using high-speed imaging, we checked that the terminal velocity $U$ was reached by the drops. Figure S1 shows snapshots of drops with volume $\Omega = 100$ µL running down a superhydrophobic incline. Consecutive snapshots are spaced by a constant $\Delta t$ so that drops in the regime of terminal velocity should travel by the same distance between each snapshot. Figure S1a shows a water drop (corresponding to the low viscosity regime in the Figure 3a of the accompanying paper) running down an incline tilted by $\alpha = 2°$, with a time step $\Delta t = 14$ ms. The red line illustrates the constant velocity $U \approx 66$ cm/s and it confirms that we reached the terminal velocity, as expected after more than 2 m of descent. The transient regime of a Galilean acceleration is expected to last $U/g\sin\alpha \approx 1.9$ s, which corresponds to a travelled distance of $U^2/2g\sin\alpha \approx 60$ cm. We also show in Figure S1b a similar experiment for a viscous drop $\eta = 110$ mPa.s. In this case, we have $\Delta t = 80$ ms and $\alpha = 8.6°$, and the measured terminal velocity is $U \approx 11$ cm/s.

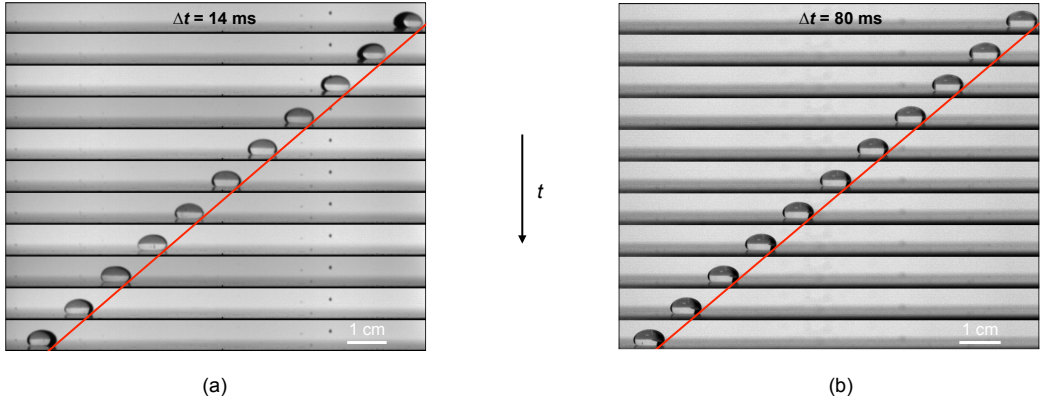

(a)  (b)

**Figure S1.** Terminal velocity of drops with volume $\Omega$ = 100 µL running down a superhydrophobic incline. **(a)** Water drop ($\eta$ = 1 mPa.s) with $\alpha$ = 2°. Snapshots are equally spaced by $\Delta t$ = 14 ms. The red line illustrates a constant velocity regime with $U \approx$ 66 cm/s. **(b)** Same experiment as in (a), for a water-glycerol drop of viscosity $\eta$ = 110 mPa.s, with $\alpha$ = 8.6° and $\Delta t$ = 80 ms. The terminal velocity is now $U \approx$ 11 cm/s.

## 2. Volume dependency of terminal velocity

As discussed in the main text, the terminal velocity $U$ results from the balance of frictions in liquid and in gas, both proportional to the drop contact area $\Sigma$, with the weight $\rho\Omega g \sin\alpha$. For large drops, the drop height is fixed by the capillary length so that the volume scales as $a\Sigma$. Hence the terminal velocity $U$ should be independent of the drop volume. In Figure S2, we plot the velocity $U$ as a function of the volume $\Omega$ for various tilt angles $\alpha$. We perform that experiment for both water-glycerol mixture of viscosity $\eta$ = 110 mPa.s (Figure S2a) and water (Figure S2b). Velocity hardly varies with the volume as expected from our model.

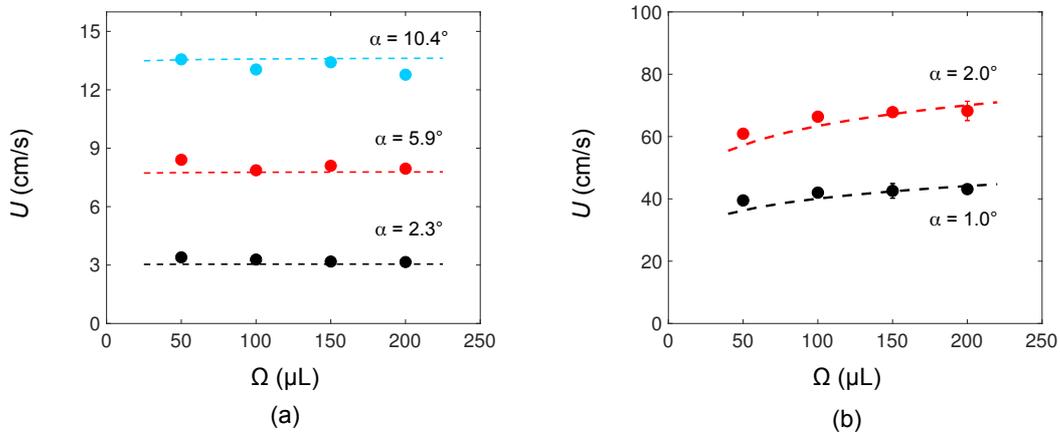

(a)  (b)

**Figure S2.** Terminal velocity $U$ of drops as a function of their volume $\Omega$ for various tilting angles $\alpha$. We present data in the viscous regime (a, $\eta$ = 110 mPa.s) and in the inviscid regime (b, $\eta$ = 1 mPa.s). Dashed lines represent Eq. 3 with $x$ = 0.85 and $y$ = 34.

## 3. Liquid and gas flow visualization

Similar experiments can be performed inside a brass cylinder coated with the same SH treatment as the incline. After placing a 100 μL water drop in the cylinder, rotation is set until reaching a prescribed velocity $U \approx 0.26$ m/s. The depth of the cylinder is 5 cm, and the track is slightly convex (curvature of 1 m$^{-1}$), which traps the liquid at the center. We perform two kinds of visualization. (1) To reveal the flow inside the drop, we add PIV tracers in the drop and illuminate a plane in the middle of the drop orthogonal to the cylinder axis of rotation. Experiments are filmed with a high-speed camera, and we extract particles motion. Using a Matlab PIV code (PIVLab [1-3]) we reconstruct the velocity profile inside the drop as shown in Figure S3a. As expected on SH materials, we observe a rolling motion inside water. (2) We similarly record high-speed film for the same drop after injecting smoke upstream. Using ImageJ, we extract the standard deviation of the recorded movie and the resulting image is displayed in Figure S3b. We clearly see a counter-rotating vortex relatively to the rolling motion, as expected from the boundary condition at the top of the rolling drop. This vortex is observed to stay attached to the drop, a signature of a skin-drag regime.

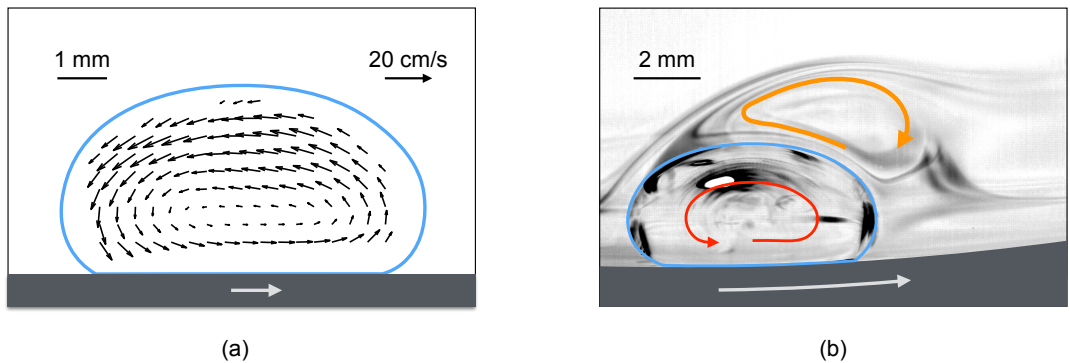

(a)          (b)

**Figure S3. (a)** PIV visualisation of the flow inside a 100 μL water drop placed in a rotating cylinder moving at 0.26 m/s. The fluid is observed to be in rotation as assumed in our model. **(b)** Smoke visualization of the gas flow around the drop (the image displayed is the standard deviation obtained from the high-speed imaging of the phenomenon). A vortex (orange arrow) rotating in the opposite rolling direction (red arrow) is observed to stay attached to the drop.

## 4. Captions for the supplementary movies

**Supplementary Movie 1.**
Water drop of volume 100 μL running down a superhydrophobic incline tilted by 2°. The movie is slowed down 40 times and the terminal velocity is 66 cm/s.

**Supplementary Movie 2.**

Water-glycerol mixture drop of volume 100 µL and viscosity η = 110 mPa.s running down a superhydrophobic incline tilted by 8.6°. The movie is slowed down 5 times and the terminal velocity is 11 cm/s.

**Supplementary Movie 3.**

Liquid and gas motion for a 100 µL drop rolling inside a superhydrophobic cylinder with a surface velocity of 0.26 m/s. The movie is slowed down 100 times.